\documentclass[twocolumn,showpacs,superscriptaddress,prl]{revtex4-2}
\usepackage{comment}
\usepackage[dvips]{graphicx}	
\usepackage{dcolumn, bm, color, amsmath}
\usepackage{amssymb}
\usepackage{tikz}
\usepackage{colortbl} 
\usepackage{array}          

\newcommand{\ffullhexagonSymbolTwo}{
  \tikz[baseline=0.07cm]{
    \draw (0.1,0.3) -- (0.25,0.225) -- (0.25,0.075) -- (0.1,0.001875) -- (-0.05,0.073125) -- (-0.05,0.225) -- cycle;
    \fill (0.1,0.3) circle (1.5pt);    
    \fill (0.1,0.001875) circle (1.5pt);   
    \fill (0.25,0.225) circle (1.5pt);     
    \fill (-0.05, 0.225) circle (1.5pt);      
    \fill (0.25,0.075) circle (1.5pt);       
    \fill (-0.05,0.073125) circle (1.5pt);         
  }
}

\begin{document}

\title{Two characteristic contributions to the superconducting state of 2$H$-NbSe$_2$}

\author{A. Alshemi}
\email{ahmed.alshemi@sljus.lu.se}
\affiliation{Division of Synchrotron Radiation Research, Department of Physics, Lund University, SE-22100 Lund, Sweden}

\author{E.~M.~Forgan}
\affiliation{School of Physics and Astronomy, University of Birmingham, Edgbaston, Birmingham, B15 2TT, UK}

\author{A. Hiess}
\affiliation{Institut Laue Langevin, 71 Avenue des Martyrs, CS 20156, 38042 Grenoble Cedex 9, France}
\affiliation{European Spallation Source ERIC, P.O. Box 176, 221 00 Lund, Sweden}

\author{R. Cubitt}
\affiliation{Institut Laue Langevin, 71 Avenue des Martyrs, CS 20156, 38042 Grenoble Cedex 9, France}

\author{J. S. White}
\affiliation{Laboratory for Neutron Scattering and Imaging, PSI Center for Neutron and Muon Sciences, Forschungsstrasse 111, 5232 Villigen PSI, Switzerland}

\author{K. Schmalzl}
\affiliation{Forschungszentrum J\"ulich GmbH, JCNS at ILL,
71 Avenue des Martyrs, 38042 Grenoble, France}

\author{E.~Blackburn}
\email{elizabeth.blackburn@sljus.lu.se}
\affiliation{Division of Synchrotron Radiation Research, Department of Physics, Lund University, SE-22100 Lund, Sweden}

\begin{abstract}
Multiband superconductivity arises when multiple electronic bands contribute to the formation of the superconducting state, allowing distinct pairing interactions and gap structures.  
Here, we present field- and temperature-dependent data on the vortex lattice structure in 2$H$-NbSe$_2$ as a contribution to the ongoing debate as to whether the defining feature of the superconductivity is the anisotropy or the multi-band nature.
The field-dependent data clearly show that there are two distinct superconducting bands, and the contribution of one of them to the vortex lattice signal is completely suppressed for magnetic fields above $\sim$ 0.8 T, well below $B\mathrm{_{c2}}$. By combining the temperature and field scans, we can deduce that there is a moderate degree of interband coupling.  From the observed temperature dependences, we find that at low field and zero temperature, the two gaps in temperature units are 13.1 $\pm$ 0.2 and 6.5 $\pm$ 0.3  K ($\Delta_{0}$ = 1.88 and 0.94 $k\mathrm{_{B}} T\mathrm{_{c}} $); the band with the larger gap gives just under two-thirds of the superfluid density.  The penetration depth extrapolated to zero field and zero temperature is 160 $\pm$ 2 nm.

\end{abstract}

\keywords{superconductivity, vortex lattice, niobium diselenide}

\date{\today}

\maketitle

Many conventional superconductors show deviations from the Bardeen-Cooper-Schrieffer (BCS) model, and one of the most studied of these is the existence of multiple gaps.  Suhl \emph{et al.}~\cite{Suhl1959_SuMW} extended the standard BCS model by considering the effects of a second energy band crossing the Fermi level.  However, experimentally it is challenging to distinguish between (strongly) anisotropic $s$-wave superconductivity with a single gap function, and true multi-gap or multi-band superconductivity \cite{Zehetmayer_2013}.

MgB$_2$ provided a very appealing model for a two-gap system, with different gaps observed for the boron 2$p$ $\sigma$ and $\pi$ bands that both open up at $T_{\mathrm{c}}$ \cite{Souma2003_SMST}.  This led to a re-evaluation of other conventional superconductors, including the subject of this Letter, 2$H$-NbSe$_2$.  This had originally been considered to be an anisotropic single-band superconductor, but the experimental evidence could also support a two-band picture \cite{HESS_1991, Boaknin2003_BTPH, RODR_2004, Fletcher2007_FCDR, Zehetmayer2010-magn}, and this question remains the subject of active debate \cite{Sanna2022_SPLR}.

\begin{figure}[t]
    \includegraphics[width=\columnwidth]{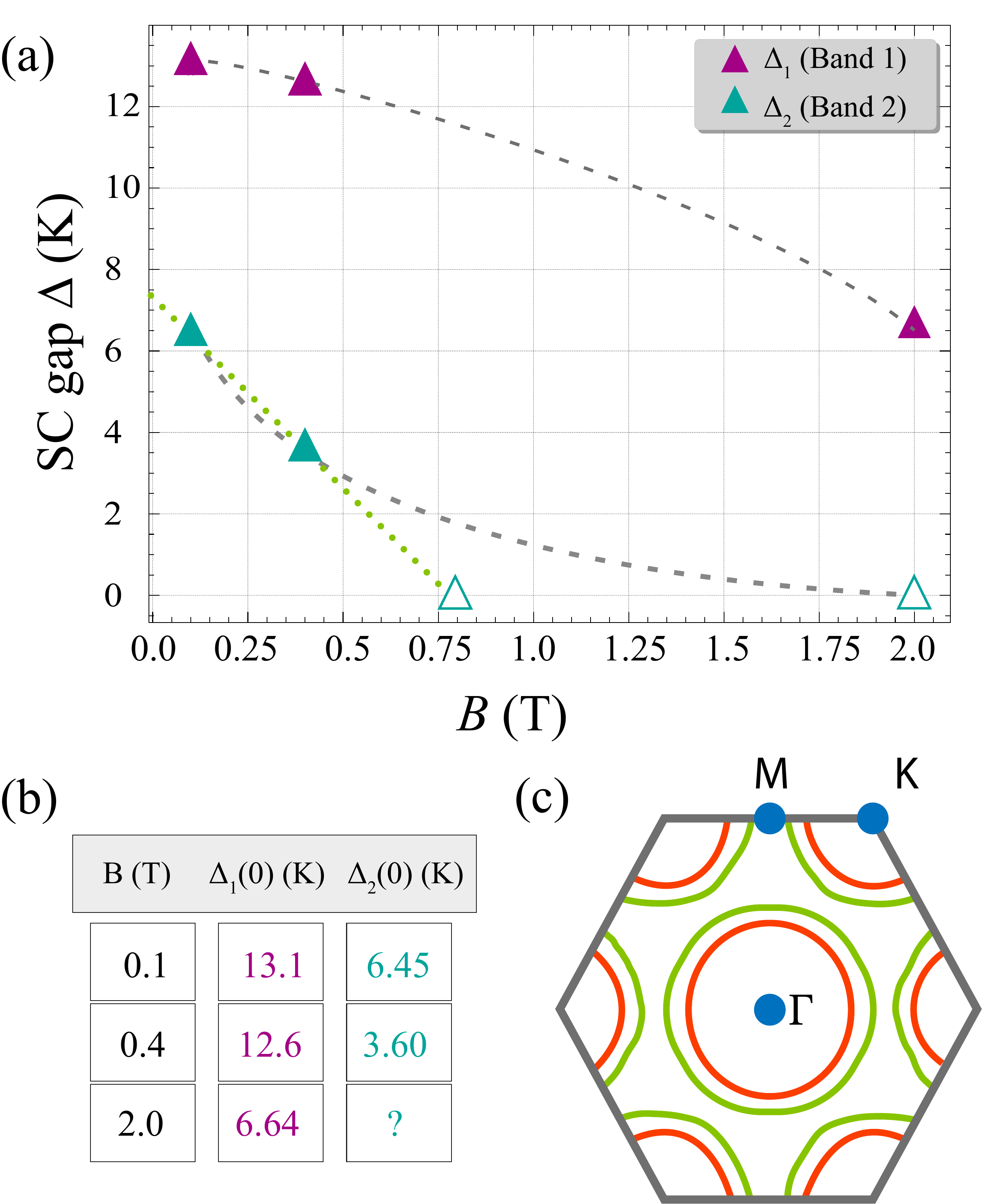}
	\caption{(a) The field dependence of the superconducting gap values for the two bands (in temperature units), as obtained from the fitting described in the main text. The open teal symbols are speculations as to where $\Delta_2$ goes to zero and do not come from fits to the data. The green dotted line is the straight line through the two values obtained for Band 2.  The gray dashed lines are guides to the eye.  (b) Table showing the numerical values in (a). The value denoted with a `?' cannot be determined from our fits as the relevant contribution is negligible due to the core cut-off factor. (c)  A sketch of the predominantly Nb components of the Fermi surface of 2$H$-NbSe$_2$ based on Ref.~\cite{Pásztor2021}. The parts shown are the cylinders centered about the $\Gamma$ and K points.  The outer cylinders (green) are anti-bonding in nature, and the inner cylinders (red) are bonding in nature.}
	\label{fig:Delta_B}
\end{figure}

The Fermi surface of 2$H$-NbSe$_2$ has been well studied \cite{Corcoran,Noat_2015}, and the main features are a set of cylinders originating from the Nb 4$d$ bands, associated with centered around the $\Gamma$ and K points in the Brillouin zone (discussed later in Fig.~\ref{fig:Delta_B}c).  There is also a smaller pancakelike sheet at the $\Gamma$ point that is associated with the Se $p$ bands.  This relatively simple picture is then dramatically complicated by the appearance of the charge density wave state, nicely illustrated in the simulations of Sanna \emph{et al.}~\cite{Sanna2022_SPLR}.

The superconductivity is generally considered to sit on the Nb 4$d$ bands.  There is some debate as to whether the main contribution is coming from the bonding or anti-bonding Nb components, or from the cylinders around the K point.
Using first principles calculations of density functional theory for superconductivity, Sanna \emph{et al.}~\cite{Sanna2022_SPLR} find that the superconducting gap varies within a given sheet and between the sheets, and use this to explain scanning tunneling spectroscopy measurements, claiming that this rules out a two-gap picture.  Other groups have measured similar spectra, and argue instead for a two-gap picture, but with a strong interband coupling.  Ref.~\cite{Noat_2015} gives a good overview of the results for both pictures drawing on results from a range of other methods, including angle-resolved photoemission spectroscopy.  We find that our results are well described by a two-gap model.

\begin{figure}[t]
    \includegraphics[width=\columnwidth]{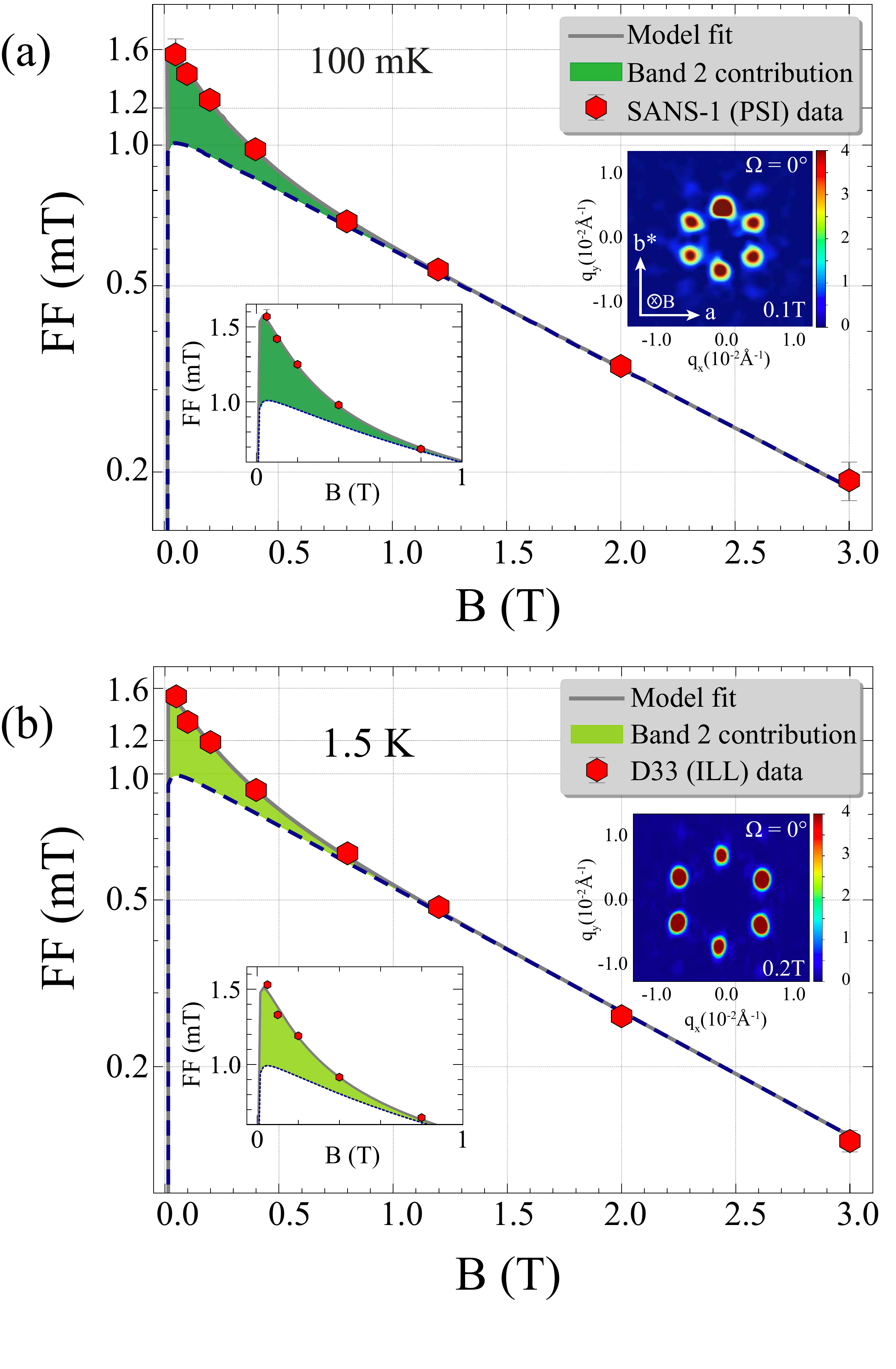}
	\caption{The field dependence of the first order vortex lattice form factor, $FF$, at 100 mK (upper panel) and 1.5 K (lower panel).  The gray lines are the fit to the data using the model described in the main text.  The dashed blue lines are the contribution from the dominant superconducting band which persists to high fields and the shaded areas indicate the deviations from the single-band picture, highlighting the presence of addition supercarriers from the second gap. The insets on the left show the data with a linear vertical scale, focusing on the low-field region. The insets on the right show example diffraction patterns measured at different fields and different instruments, showing that the alignment of the vortex lattice is unchanged. When the experimental conditions are the same, the integrated intensity obtained from the two instruments is essentially the same (see End Matter)}. 
	\label{fig:FF_Bdep}
\end{figure}

 Boaknin \emph{et al.}~\cite{Boaknin2003_BTPH} argued that thermal conductivity and heat capacity measurements as a function of field could be explained by a two-gap/two-band explanation, in part because of the close resemblance to the field dependences observed in MgB$_2$.  In that material, the magnetic field suppresses the contribution of the $\pi$ band well below the upper critical field $B_{c2}$, as illustrated by Cubitt \emph{et al.}~\cite{Cubitt-MgB2} in their neutron diffraction study of the vortex lattice in MgB$_2$. Using the same technique, we show that in 2$H$-NbSe$_2$, there is an almost identical suppression of one of the contributions to the overall superconducting state (see Fig.~\ref{fig:FF_Bdep}). We argue that this arises due to two different core-sizes arising from the existence of two bands with different gap values.  
This cannot easily be explained in a purely anisotropic $s$-wave case, and requires two bands.

The vortex lattice (VL) in this material has previously been observed by neutron diffraction \cite{Gammel1994, Yaron1994_YGHK, Marziali2015}.~These works examined the vortex lattice structure and perfection; here we concentrate on the field- and temperature-dependence of the form factor. Gammel \emph{et al.}~\cite{Gammel1994}~saw a perfect hexagonal lattice when the field is applied parallel to \textbf{c}. This means that the superconducting parameters are essentially isotropic within the $ab$ plane.  Only one hexagonal domain is observed, of the form \ffullhexagonSymbolTwo, with the spots along $\bf{b^*}$-type axes. Generally, non-local effects may lock the vortex lattice orientation in this way, although Gammel \emph{et al.}~\cite{Gammel1994} argue that here it is linked to the charge density wave state that coexists with superconductivity in NbSe$_2$, as its propagation direction is also along the $\bf{b^*}$-type axes. Scanning tunneling microscopy studies show that the vortex cores themselves possess a six-fold symmetry \cite{Hess-star-shape}, and this has also been advanced as an explanation for the existence of a particular VL orientation \cite{Ganguli_2016}.  Note that our sample is rotated about the {\bf c} axis by 90$^{\circ}$ relative to Gammel \emph{et al.}~\cite{Gammel1994}.

The ordered arrangement of flux lines in the superconducting state generates a periodic modulation of magnetic field that scatters the neutrons.  In a single-band Type-II superconductor, the intensities of the resulting Bragg reflections are completely determined by the magnetic field, the London penetration depth $\lambda$ and the superconducting coherence length $\xi$ \cite{Eskildsen2011_EsFK}. We measure the form factor, which is the Fourier component of the spatial variation of field inside the vortex lattice at the momentum transfer $q$ of a given reflection, and is usually well described at low temperatures by the London model, modified with a core cut-off term \cite{Yaouanc_1997} such that
\begin{equation}
F(q)=\frac{ B \exp \left(-c q^2 \xi^2\right)}{1+q^2 \lambda^2}.
\label{LM}
\end{equation}
The magnetic field applied is given by $B$, which is proportional to $q^2$ as each vortex carries one flux quantum.  The constant $c$ is the core cut-off parameter; we use $c$ = 0.44 here \cite{Campillo2022_YBCO}.  This means that $\xi$ represents the effective vortex core size, which is the coherence length in a single-band superconductor.

The experimental data shown here were collected at the D33 instrument at the Institut Laue-Langevin \cite{Dewhurst:ks5488,ILLdata} and the SANS-I instrument at the Swiss Neutron Spallation Source SINQ, using the same single crystal, described in the End Matter. This is the same sample as studied by Schmalzl \emph{et al.}~\cite{Schmalzl_INS}; the $\bf{c}$-axis was aligned parallel to the applied magnetic field, which was in turn almost parallel to the incident neutron wavevector.  Details of the instrument settings and preparation of the vortex lattice are given in the End Matter.

The field dependence of the form factor at 100 mK and 1.5 K is shown in Figure \ref{fig:FF_Bdep}, with the form factor plotted on a log scale.  As the Bragg reflections are all in the same orientation with respect to the crystal axes, we do not have to worry about anisotropy in the penetration depth or coherence length, even though this is in principle a highly anisotropic 2D superconductor. The blue dashed line in Figure \ref{fig:FF_Bdep} illustrates the behavior described in Eq.~\ref{LM} - the steeper the slope, the greater the value of $\xi$, while the zero field value of $\lambda$ would be obtained by linear extrapolation of this line to the vertical axis.  

It is clear from the data that this is not the case; at lower fields, there is a deviation from this simple description.  To address this, we start with Prozorov and Giannetta's model \cite{Prozorov2006_PrGi} for the superfluid density of a clean two-gap superconductor. In the following expression, the densities are normalized to their values at $T = 0$:
\begin{equation}
\rho_{\text{s}}(T) = x \, \rho_{\text{1}}(T) + (1 - x) \, \rho_{\text{2}}(T)
\end{equation}
where $x$ represents the relative contribution of Band 1 to the overall superfluid density and can be calculated by integrating the Fermi velocity over the Fermi surface.  $x$ should be independent of magnetic field and temperature.  $\rho_{1,2}$ represents the temperature dependences arising from the gaps associated with the different Fermi surfaces. This total normalized superfluid density is directly related to the penetration depth, $\rho_s = \lambda_0^2/\lambda^2(T)$, where $\lambda_0$ is the penetration depth at 0 K.  

We can insert this into Eq.~\ref{LM}, but here it is helpful to apply the approximation that $q^2\lambda^2 \gg 1$. At 0.05 T, $q^2\lambda^2 \sim$ 30, so this approximation is good for all measured fields.  When we apply this approximation, it is then easy to include separate core sizes 
associated with each band \cite{Silaev-Babaev,Ichioka_2017,ChenShanenko-healinglengths}, such that we can rewrite Eq.~\ref{LM} as 
\begin{equation}
F(q)= \frac{B}{q^2 \lambda_0^2} \Big( x \rho_1(T) \mathrm{e}^{-c q^2 \xi_1^2} + (1-x) \rho_2(T) \mathrm{e}^{-c q^2 \xi_2^2}  \Big).
\label{LM-adapted}
\end{equation}
Since the vortex lattice is perfectly hexagonal, $q^2 \propto B$, this can be rewritten into the functional form used by Cubitt \emph{et al.}~for MgB$_2$ \cite{Cubitt-MgB2} as
\begin{equation}
F(q)= \frac{B \mathrm{e}^{-c q^2 \xi_1^2}}{q^2 \lambda_0^2} \Big( x \rho_1(T) + (1-x) \rho_2(T) \mathrm{e}^{-B/B^*}  \Big).
\label{LM-cubitt}
\end{equation}
where $B^* = \sqrt{3}\Phi_0 / 8 \pi^2 c (\xi_2^2 - \xi_1^2).$

To fit our data to this model, we use the method given in the End Matter to calculate the normalized superfluid density and the core size for each of the bands, labeled $i$, as functions of temperature.

\begin{figure}[th]
    \includegraphics[width=0.9\columnwidth]{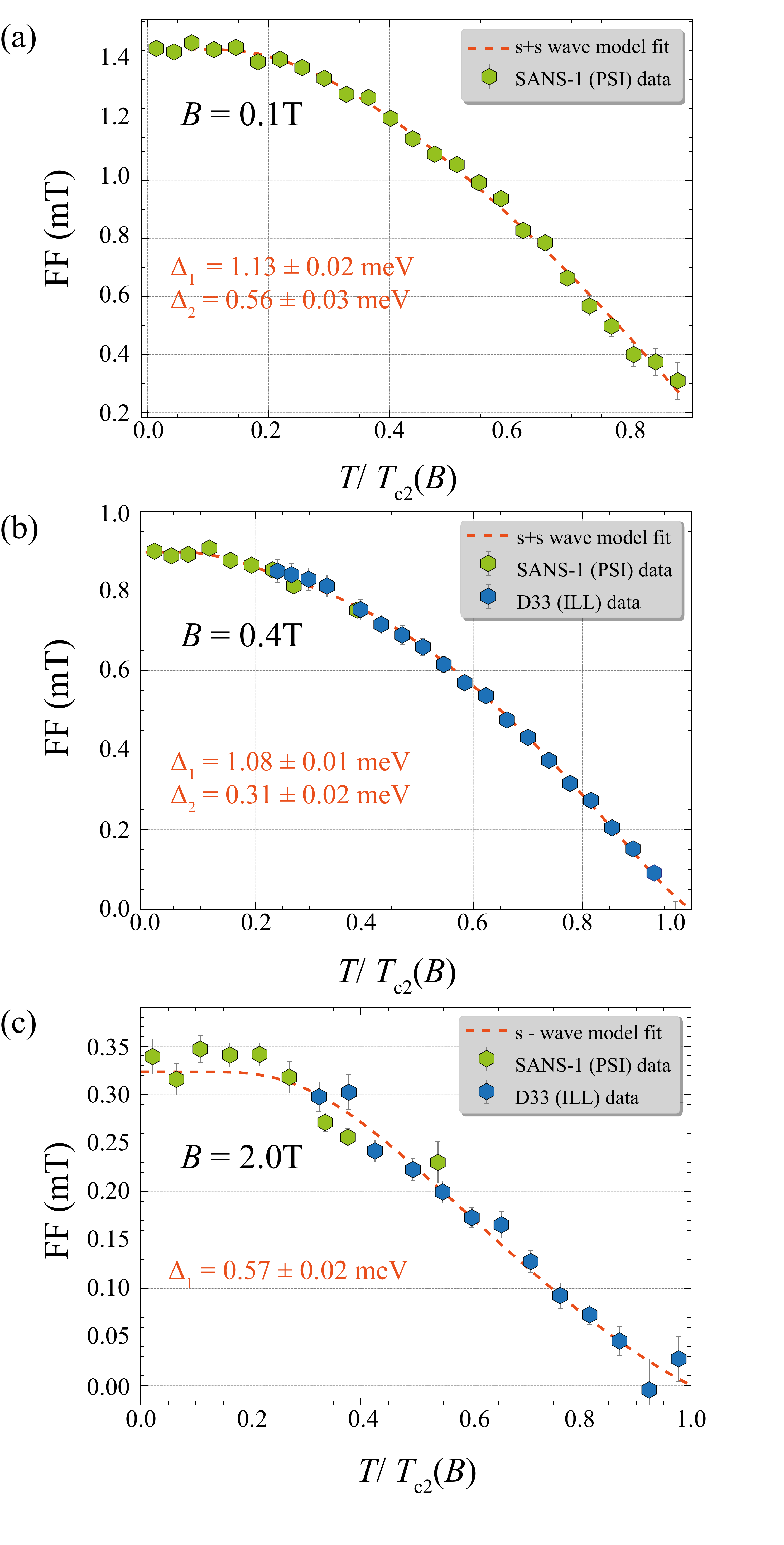}
	\caption{The temperature dependence of the first order vortex lattice form factor, $FF$, measured at $B$ = 0.1, 0.4 and 2.0 T.  The temperature values are normalized to the critical temperatures ($T_{\mathrm{c2}}$) at the applied magnetic field.  $T_{\mathrm{c2}}$(0.1 T) = 6.85 K, $T_{\mathrm{c2}}$(0.4 T) = 6.46 K and $T_{\mathrm{c2}}$(2.0 T) = 4.63 K.}
	\label{fig:FF_temp}
\end{figure}

From the field-dependent data in Fig.~\ref{fig:FF_Bdep} we can establish values for $x$ and $\xi_i$.  The values of $\xi_i$ at 100 mK are used as our best estimate for the zero temperature values, together with an average of the values for $x$ (= 0.61). By fitting the temperature-dependent data (Fig.~\ref{fig:FF_temp}), values for the energy gaps $\Delta_i$ are obtained.  After iteration to arrive at the optimal solution, we obtain the penetration depth and coherence length values given in Table \ref{tab:fit_parameters}.

\begin{table}[h]
    \centering
    \begin{tabular}{ccccc}
        \hline
         & \textbf{$\lambda$ (nm)} & \textbf{$\xi_{1}$ (nm)} & \textbf{$\xi_{2}$ (nm)} \\
        \hline
        $B$ scan @ 0.1 K & 160 $\pm$ 3 &  7.8 $\pm$ 0.2 & 21 $\pm$ 2  \\
        $B$ scan @ 1.5 K  & 161 $\pm$ 3 &  8.4 $\pm$ 0.2 & 21 $\pm$ 2  \\
        $T$ scan @ 0.1 T  & 160.7 $\pm$ 0.4 &  fixed & fixed \\
        $T$ scan @ 0.4 T & 162.9 $\pm$ 0.5 &  fixed & fixed \\
       $T$ scan @ 2.0 T & 162 $\pm$ 2  &  fixed & N/A  \\
        \hline
    \end{tabular}
    \caption{The extrapolated penetration depths and core sizes at either 0 T (for $T$ scans) or 0 K (for $B$ scans) obtained from the fits to the field- and temperature-dependence of the vortex lattice form factor described in the main text.  The fixed $\xi_i$ values are set to the values from the $B$ scan at 100 mK.}
    \label{tab:fit_parameters}
\end{table}

The value for $\xi_1$ expected by calculation from $B_{c2}$ is 8.5 nm, close to the result for Band 1 from the fit of the field dependence at 1.5 K.  The change to 7.8 nm at 0.1 K is assigned to Kramer-Pesch shrinkage of the vortex cores \cite{KPshrinkage}, and matches the value obtained by Hess \emph{et al.} at 0.3 K \cite{Hess-star-shape}. This effect is small, as we expect given the level of cleanliness for our sample (see End Matter) and so we do not consider this effect when fitting our temperature-dependence data.  For Band 2, we find $\xi_2$ = 21 nm.  From this we calculate a value for the $B^*$ in Eq.~\ref{LM-cubitt} of $B^*$ = 0.28 T at 1.5 K and 0.27 T at 0.1 K.

From the fitted values of the penetration depth, we see that $\lambda$ at zero temperature and zero field is $\sim$ 160 nm.  This includes contributions from both bands. For the temperature dependence at $B$ = 2.0 T, only one band contributes as the other is suppressed by the core cut-off effects ($\xi_2 \sim$ 21 nm). We find that the contribution to the penetration depth from this band is $\sim$ 200 nm.  These numbers match with field-dependent $\kappa$ values from reversible magnetization measurements \cite{Zehetmayer2010-magn}, as well as with a calculation following Kogan \cite{Kogan_lambda_2020} (see End Matter). The band that persists to high fields (Band 1) has $\xi_1 \sim$ 8 nm.  
We have tested that the assignment of the $\xi_i$ values to the different bands cannot be switched; if they are, impossibly large values for $\Delta_1(0)$ result.  Based on these fits, we are able to obtain values for $\Delta_i(0)$ for the two bands as a function of field in Fig.~\ref{fig:Delta_B}b.  We cannot assign a value for $\Delta_2(0)$ at 2.0 T, because the contribution from this band is negligible due to the core cut-off factor.

Two length scales have also been used as a tool to explain the thermal conductivity at very low temperatures \cite{Boaknin2003_BTPH}.  They explain this in terms of overlap of cores permitting quasiparticle delocalization.  The ratio of our $\xi_{2,1}$ values (2.7) is similar to their value of $\sqrt{H_{\mathrm{c2}}/H^*}$ $\sim$ 3.    

It has been argued \cite{Kusunose2002_KuRS} that this delocalization occurs if the intrinsic coherence lengths for the individual superconducting sheets are different, and if one of them is comparable to the electronic mean free path $\ell$ over the entirety of that Fermi surface.  This delocalization then results in a shrinking of the vortex core when one of the sheets no longer contributes to the effective core-size.  When considering this, the relative cleanliness of the superconductor is important.  We find the ratio of $\ell / \xi_0$ to be $\sim$ 3 (see End Matter).  This also supports our contention that the Kramer-Pesch shrinkage is not strong.
However, in a muon spin rotation study by Callaghan \emph{et al.}~\cite{Callaghan2005_CLKS}, a single-band approach was used with both $\xi$ and $\lambda$ treated as continuously varying in magnetic field.  This gives the same qualitative behavior as both our results and those in Ref.~\onlinecite{Boaknin2003_BTPH}.

Our model, with two different core sizes associated with the two bands gives a good account of the data and we now compare it with theoretical expectations.  Ichioka \emph{et al.}~\cite{Ichioka_2017} have considered a clean two-band $s$-wave superconductor with cylindrical Fermi surfaces with differing Fermi velocities, which should be a good proxy for NbSe$_2$, and calculated the variation in $\Delta$ and $\xi$ as a function of field and temperature for both strong and weak interband coupling.  They find that when two bands are strongly coupled, the core sizes lock at a single value, but when the bands are weakly coupled, two different lengths exist. So are the two bands in NbSe$_2$ weakly coupled? For that case, Ichioka \emph{et al.}~\cite{Ichioka_2017} find that the lower gap would not have a BCS temperature-dependence and would drop almost to zero well below $T_{\mathrm{c}}$. To test this possibility, we have introduced into the temperature-dependence a $T^*$ analogous to $B^*$. However, our data cannot be fitted with a $T^* < T_{\mathrm{c}}$, so we rule out the weak coupling scenario, in agreement with other authors \cite{Noat_2015}.

Chen and Shanenko~\cite{ChenShanenko-healinglengths} differ from Ref.~\cite{Ichioka_2017} in that they calculate Cooper pair wavefunctions and find that the Cooper pair density varies on a different length-scale to $\Delta$ near a (single) vortex core. Using MgB$_2$ as their reference, they find two distinct length-scales for the Cooper pair density 
at low temperatures, even with strong interband coupling. As they point out, there are many ways to define the core sizes.  In our case, we take it to be given by the core cut-off factor. Qualitatively, their results are closer to our model, though the two lengths become equal at $T_{\mathrm{c}}$, where our temperature-dependent data have less sensitivity (see End Matter). A single length-scale for the superconductivity at $T_{\textrm c}$ is the expected result from two-band Ginzburg-Landau theory~\cite{Silaev-Babaev}.

We now turn to the field dependence of $\Delta_i$. Our low-field gap values (Fig.~\ref{fig:Delta_B}a,b) and relative weighting match well with those obtained from tunnel diode oscillator measurements of the penetration depth \cite{Fletcher2007_FCDR}, which measure a similar quantity.  They are slightly lower than the values obtained at zero field from heat capacity, photoemission and scanning tunnelling spectroscopy (summarized in Ref.~\cite{Noat_2015}). In Fig. \ref{fig:Delta_B}a, $\Delta_1$ behaves as expected for the average energy gap in the mixed state. For $\Delta_2$, the two lines represent two scenarios.  Following the straight (green dotted) line through the two gap values obtained for Band 2, the gap would close at 0.8 T ($\sim$ 3$B^*$).  In the other (gray dashed) line, $\Delta_2$ decreases to near zero at 2 T. Our actual results up to 0.4 T lie between the predictions of Ref.~\cite{Ichioka_2017} in the weak and strong coupling cases, while Chen and Shanenko~\cite{ChenShanenko-healinglengths} do not calculate the field dependence of $\Delta_i$. We conclude that neither theory is fully consistent with our model and that there is moderate interband coupling in NbSe$_2$.

We are not directly sensitive to the charge density wave response in the measurements reported here.  However, we note that the suppression of the second band due to core overlap at fields above $\sim$ 0.8 T correlates with the observation by Raman scattering ($A_{1g})$ of a transfer of intensity from an enhanced superconducting pair-breaking peak to the soft phonon mode of the charge-density-wave state \cite{Raman-field-dep}. 

To summarize, by considering the ensemble of data measured on the vortex lattice with the field applied parallel to the $\mathbf{c}$ axis in NbSe$_2$, we find clear evidence for two contributions to the vortex core size in the field-dependence of the vortex lattice signal.  From the size of these characteristic lengths we conclude that a significant part of the vortex lattice form factor disappears at fields above $\sim$ 0.8 T 
due to the core overlap associated with the larger length-scale (21 nm).  We conclude that NbSe$_2$ therefore has two separate contributions to the superconducting response.  Within the constraints of our model, we extract the field dependence for the values of the superconducting gap for these two components and also deduce that the two bands are moderately strongly coupled.  
This combination of field- and temperature-dependent data could be successfully applied to other candidate multi-band superconductors, such as MgB$_2$, to test if the interband coupling is strong or weak.

\section*{Acknowledgements}
AA and EB acknowledge support from the Crafoord Foundation (no.~20190930) and the Swedish Research Council under Project No.~2021-06157.  AA and EMF thank the Institut Laue-Langevin for travel support to attend the experiment.  The authors gratefully acknowledge the Institut Laue-Langevin and the Swiss Neutron Spallation Source SINQ for the allocated beamtime.  We thank Matthew Coak for carrying out magnetization and resistance measurements on the sample and Dmytro Orlov for assistance with the energy dispersive X-ray measurements.  We thank Egor Babaev, Alistair Cameron and Emma Campillo for helpful discussions about the manuscript.

\section*{End Matter}
\noindent \textbf{Experimental Details:} 2$H$-NbSe$_{2}$  samples were grown using the chemical vapor transport technique, yielding high-quality single-crystals with optically flat surfaces on the macroscopic scale.  The sample quality was confirmed by magnetization and resistance measurements, wherein a sharp jump centered at the superconducting transition temperature $T_{\rm c}$ = 6.95 K is seen (as measured after the neutron scattering experiments).  This value of $T_c$ is used to establish the relevant critical temperatures used in the fitting process described in the main text, following the phase diagram measured by Cho \emph{et al.}~\cite{Cho_2022}.

The 2$H$-NbSe$_{2}$ polytype is hexagonal, with space group $P6_3/mmc$ and the room temperature lattice parameters are $a = b \sim 3.44$ \AA~ and $c \sim 12.55$ \AA.  The samples grow as plates with the \textbf{c} axis normal to the surface. A hexagonal single crystal of mass 93.8 mg and size 5 x 7 x 1 mm$^3$ was mounted.  The vertical axis was $\mathbf{b^*}$.  Energy dispersive X-ray spectroscopy points to a possible small excess of selenium.

To create the vortex lattices, a horizontal magnetic field was applied parallel to the sample $\mathbf{c}$ axis (approximately parallel to the incoming neutron beam).  At D33, the field was applied in the normal state and then the sample was cooled to the base temperature of 1.5~K while slowly oscillating the external field by $\pm 1\%$. 
At SANS-I, the 11 T horizontal magnet MA11 was used with a dilution insert.  This gave a much lower base temperature of 100 mK, but made it very time-consuming to heat The sample above $T_{\mathrm{c}}$. Given the limited experimental time available, the lattice was formed by first cooling to the desired temperature in a magnetic field greater than $H_{c2}$ (5.5 T) and then reducing the field  to the target value to measure the VL.  

Several combinations of neutron wavelength, collimation and detector distance (W/C/D) were used at the two instruments to be able to measure at different fields.  The wavelength spread was $\Delta\lambda_{n}/\lambda_{n}$ = 10\% full-width half-maximum (FWHM). A circular aperture of diameter 6 mm was placed close to the sample in all experiments. Different instrument settings were used to capture the full range of fields used (see Table \ref{tab:inst_params}).  

\begin{table}[h]
    \centering
    \begin{tabular}{ccccc}
        \hline
        \textbf{Instrument } & \textbf{$\lambda_{n}$} & \textbf{Collimation} & \textbf{Det.~Dist.} & \textbf{Fields} \\
        & (\AA) & (m) & (m) & (T) \\
        \hline
        D33 & 10 &  12.8 & 10 & 0.2 - 1.2\\
        D33 & 10 &  12.8 & 5 &  2.0 - 3.0\\
        D33 & 6 &  12.8 & 12 & 0.4 - 2.0\\
        D33 & 10 &  12.8 & 12 & 0.05 - 0.1 \\
        SANS-I & 6  & 18 & 17 & 0.1 - 3.0 \\
        SANS-I & 10  & 18 & 18.2 & 0.05 - 0.4\\
      
        \hline
    \end{tabular}
    \caption{Different instrumental settings used for both field and temperature dependent measurements. The relevant magnetic field ranges are specified.}
    \label{tab:inst_params}
\end{table}
To obtain the integrated intensities of the Bragg reflections associated with the vortex lattice, the sample and magnet were rocked together about the axes perpendicular to the beam direction.  At D33, the background measurements were collected in the normal state, at 8.5 K ($> T_\mathrm{c}$).  On SANS-I, they were collected at 5.5 T ($> B_{c2}$) for a given temperature.
The background was then subtracted from the foreground, leaving only the vortex lattice signal. To ensure consistency between measurements taken with the two different instruments, a correction factor of 1.05 was applied to the PSI data to align it with the results obtained on D33 in regions where overlap measurements were carried out. The initial data reduction was done using the software program GRASP \cite{Dewhurst}.  The form factors were then calculated using the Christen formula \cite{Christen1977}.  Processed experimental data are available upon request to the corresponding authors. The raw data from the experiment at the Institut Laue-Langevin are available in Ref.~\cite{ILLdata}.
\\

\noindent \textbf{Superfluid density in a two-band model:}
We follow Prozorov and Giannetta \cite{Prozorov2006_PrGi}, who give the normalized superfluid density of band $i$ as:

\begin{equation}
\rho_{i}(T) = 1 - \frac{1}{2 T}  \int_0^\infty \cosh^{-2} \left( \frac{\sqrt{\epsilon^2 + \Delta_i^2(T)}}{2 k_B T} \right) d\epsilon
\end{equation}
for an isotropic $s$-wave band on a cylindrical Fermi surface, where $i$ is the band label. $\sqrt{\epsilon^2 + \Delta_i^2(T)}$ defines the excitation energy spectrum, and we model the superconducting gap functions $\Delta_i(T)$ as 
\begin{equation}
\Delta_i(T) = \Delta_i(0)\tanh\left(1.78\sqrt{\frac{T_c}{T} - 1}\right),
\end{equation}
where \(\Delta_i(0)\) is the magnitude of each gap at $T$ = 0 K.  This gives a good approximation to the Bardeen-Cooper-Schrieffer temperature dependence.  We also use this expression to give the temperature dependence of $\xi(T)$, which we take as proportional to $1/\Delta(T)$:
\begin{equation}
\xi_{i}(T) = \xi_{i}(0)\left[\tanh\left(1.78\left(\frac{T_c}{T} - 1\right)\right)\right]^{-1}.
\label{eq:xi}
\end{equation}

In our fits we assume the above temperature dependence for $\xi_i$.  We have tested changing this by adding in a temperature control parameter so that $\xi_2$ tends linearly to $\xi_1$ at $T_{\mathrm{c}}$, to approximate the results of Chen and Shanenko \cite{ChenShanenko-healinglengths}, and to give a single length-scale at $T_{\rm c}$~\cite{Silaev-Babaev}. We obtain a similar quality of fit to those shown in Fig.~\ref{fig:FF_temp}. The values of $\lambda$ and $\Delta_1$ are unchanged, but $\Delta_2$ decreases by $\sim$ 20\%. 

\noindent \textbf{Is it a clean superconductor?}
On the assumption that the material is a clean superconductor, and averaging the properties of the two bands, we can estimate the ratio of mean free path to coherence length in our sample of NbSe$_{\mathrm{2}}$. This is only an approximate calculation, but the result shows that the material is a fairly clean superconductor.

For a clean single-band superconductor,

\begin{equation}
  \frac{ne^2}{m^*} = \frac{1}{\mu_0 \lambda^2} = \frac{\sigma}{\tau},
\end{equation}
 
where the conductivity $\sigma$ is related to the average impurity scattering time $\tau$ via the Drude relationship. Hence we can estimate $\tau$ if we have the conductivity as well as $\lambda$.  From our fits, we take $\lambda$ = 160 nm.  Taking the resistivity at 300 K to be \(1.1 \times 10^{-6} \, \Omega \, \mathrm{m}\) \cite{resistivity-RT}, and given that the residual resistance ratio (RRR) of our sample is $\sim 20$, we find that $\sigma = 1.8 \times 10^{7} \, \Omega^{-1} \, \mathrm{m}^{-1}$.  This gives a scattering time $\tau = 5.9 \times 10^{-13}$ s.

The ratio of the mean free path $\ell$ to $\xi_0$ determines whether the superconductor is clean or dirty. Now: 
\begin{equation}
  \frac{\ell}{\xi_0} = \frac{v_F \tau}{\hbar v_F / \pi \Delta_0}.
\end{equation}
The Fermi velocity $v_F$ cancels on the right-hand side, so its value is not required, and the cleanliness ratio depends only on $\tau$ and $\Delta_0$. Using the BCS value of $\Delta_0 / k_B T_c$ (1.764) we obtain a ratio of $\ell/\xi_0$ = 3. 

This confirms our initial assumption that it is not a dirty superconductor; however, it is not particularly clean.  If it were dirty, this method would not be applicable, as $\lambda$ would be increased above the value given by $n/m^*$. 
We can confirm this by making the contrary assumption that the material is in the dirty limit, using the Abrikosov-Gorkov formula \cite{AbrikosovGorkov}
\begin{equation}
  \frac{1}{\mu_0 \lambda^2} = \frac{\pi \Delta_0 \sigma}{ \hbar},
\end{equation}
recently derived more accessibly in Ref.~\cite{Dutta_2022}. Using the same values of $\Delta_0$ and $\sigma$ as before, we obtain $\lambda = 93$~nm, which disagrees with our experimental value and hence rules out the dirty limit.

We have found that our material is not super-clean, which means that the Kramer-Pesch effect~\cite{KPshrinkage} is not strong. Indeed, we obtain \(\xi(1.5 \, \mathrm{K}) = 8.4 \pm 0.2 \, \mathrm{nm}\), while \(\xi(100 \, \mathrm{mK}) = 7.8 \pm 0.2 \, \mathrm{nm}\). This difference is not large, and so in our fits to the temperature-dependence, we ignore the Kramer-Pesch effect. 
\\

\noindent \textbf{Penetration depth:} The model developed by Kogan \emph{et al.}~\cite{Kogan_lambda_2020} for the zero temperature penetration depth is 
\begin{equation}
\lambda^{2}(0) \approx \left| \left(\frac{dH_{c2}}{dT}\right)_{T_c} \right| \frac{1}{T_c \gamma},
 \end{equation}
where $\gamma$ is the specific heat coefficient per unit volume.  Using this model we have calculated $\lambda$(0) = 199 nm for 2$H$-NbSe$_2$, taking as inputs  $T_{\rm c} = 6.95~K$, $(dH^c_{c2} / dT)_{T=T_c} \approx - 0.69 \times 10^4 \ \text{Oe/K}$ ~\cite{Cho_2022, Sanchez1995}, and ~ $\gamma = 0.52 \times 10^4 \ \text{erg/cm}^3 \text{K}^2$~\cite{Bevolo1974}.  This is comparable with our high field $\lambda \sim 200 $~nm at 100 mK \& 1.5 K extracted from the fits.  This model was developed for single-band $s$-wave superconductors, but should still apply if the order parameter is constant over a Fermi surface of any given shape. It has been shown to be relevant for MgB$_2$ \cite{Kogan_lambda_2020} and NbS$_2$ \cite{Alshemi2024}.
\\


\end{document}